\pdfoutput=1
\documentclass[preprint,showpacs,preprintnumbers,amsmath,amssymb,floatfix,endfloats*]{revtex4}

\usepackage{graphicx}
\usepackage{dcolumn}
\usepackage{bm}
\usepackage{amsfonts}

\DeclareMathAlphabet{\mathsfsl}{OT1}{cmr}{bx}{it}
\begin{document}
\title{Structural transformations in porous glasses under mechanical loading. II. Compression}
\author{Nikolai V. Priezjev$^{1,2}$ and Maxim A. Makeev$^{3}$}
\affiliation{$^{1}$Department of Mechanical and Materials
Engineering, Wright State University, Dayton, OH 45435}
\affiliation{$^{2}$National Research University Higher School of
Economics, Moscow 101000, Russia}
\affiliation{$^{3}$Department of Chemistry, University of
Missouri-Columbia, Columbia, MO 65211}
\date{\today}
\begin{abstract}

The role of porous structure and glass density in response to
compressive deformation of amorphous materials is investigated via
molecular dynamics simulations.  The disordered, porous structures
were prepared by quenching a high-temperature binary mixture below
the glass transition into the phase coexistence region.  With
decreasing average glass density, the pore morphology in quiescent
samples varies from a random distribution of compact voids to a
porous network embedded in a continuous glass phase. We find that
during compressive loading at constant volume, the porous structure
is linearly transformed in the elastic regime and the elastic
modulus follows a power-law increase as a function of the average
glass density. Upon further compression, pores deform significantly
and coalesce into large voids leading to formation of domains with
nearly homogeneous glass phase, which provides an enhanced
resistance to deformation at high strain.

\end{abstract}



\maketitle

\section{Introduction}

The prediction of the mechanical response of disordered solids is
important for a number of industrial applications, and, at the same
time, it poses a challenging fundamental
problem~\cite{Barrat17,Branicio17}. It is well recognized by now
that deformation and flow of bulk metallic glasses occur via rapid
localized rearrangements of atoms that induce strongly anisotropic
stress redistribution over long distances~\cite{Argon79,Spaepen77}.
At the mesoscopic level, this process can be described by
elastoplastic models, where the system is coarse-grained into
interacting elements that obey a set of rules including linear
elastic response, local yield criterion, stress propagation, and
recovery~\cite{Barrat17}.  Interestingly, atomistic simulations
revealed that both the yield and flow stresses of metallic
glasses~\cite{Schuh03} and nanocrystalline metals~\cite{Schuh04} are
higher in compression than in tension. More recently, it was shown
that several factors affect deformation and failure of cellular
metallic glasses under compression; namely, the cell size controls
the transition from localized to homogeneous plastic deformation,
while the cell shape, e.g., circular versus hexagonal, might change
the strength and energy absorption capacity due to variation in
stress concentration at the cell surface~\cite{Sha16}. Nevertheless,
a complete understanding of the elastic response and yield in
homogeneous and porous metallic glasses is yet to be achieved.

\vskip 0.05in

A number of recent experimental studies have reported the results of
uniaxial compression tests performed on metallic glass
pillars~\cite{Ma08,Spaepen08,Hosson10,Greer11,Flores11,Hosson11,Jang12}.
Most importantly, it was found that when the sample size is
decreased down to the submicron dimensions, the deformation mode
changes from shear band propagation to homogeneous plastic flow,
which can be attributed to the existence of a critical strained
volume required for the formation of a shear band~\cite{Spaepen08}.
The observed behavior can be rationalized by realizing that
collectivity of flow defects, or shear transformation zones, toward
localization is suppressed in sufficiently small systems, and the
enhanced ductility corresponds to a large number of weakly
correlated shear transformations~\cite{Ma08}. It was also shown that
during compression of micron-scale amorphous silica pillars, the
plastic deformation is accompanied with a periodic array of radial
cracks at the top of the pillars, which results in some case in
splitting into two parts upon unloading~\cite{Barthel12}. However,
despite significant efforts, the correlation between ductility,
fracture, and strength of amorphous materials as well as the
dependence on preparation history and loading conditions remain not
fully understood to date.

\vskip 0.05in

The microscopic mechanisms of the glass-gas phase separation
kinetics at constant volume were recently studied using molecular
dynamics simulations~\cite{Kob11,Kob14}. Following a rapid quench
below the glass transition temperature, a simple glass-forming
system was found to gradually transform into an amorphous solid with
a porous structure whose properties depend strongly on the average
glass density and temperature~\cite{Kob11,Kob14}.  Interestingly, it
was shown that the pore-size distribution functions obey a single
scaling relation at small length scales for systems with high
porosity, while the local density of the solid phase remains
relatively insensitive to the total pore volume~\cite{Makeev17}.
Later studies have examined the dynamic response of porous glasses
subjected to either steady shear~\cite{Priezjev17} or
tensile~\cite{Priezjev18} deformation. In both cases it was found
that the porous structure becomes significantly modified due to pore
redistribution and coalescence into large voids upon increasing
strain~\cite{Priezjev17,Priezjev18}. The analysis of local density
profiles during tensile loading showed that necking develops in the
low-density regions leading to an extended plastic strain and
ultimate breaking of the material~\cite{Priezjev18}.

\vskip 0.05in


In our recent study~\cite{Priezjev17}, we discussed the theoretical
models, developed to describe the elastic moduli of porous materials
and compared our simulation results on shear deformation of porous
glasses with analytical predictions. We found that the simulated
modulus dependence on density can not be described using a single
theory. However, the data can be fitted in the limits of low and
high porosities using different approaches (see~\cite{Priezjev17}
and references therein). In the limit of large porosities, the
percolation theory was found to adequately describe the simulation
data. In the limit of low porosities, a model, based on the Eshelby
approach to the problem of embedded inclusions, can be utilized. The
general conclusion of the study~\cite{Priezjev17} is that elastic
response properties of porous materials are strongly dependent on
the particular realization of pore-size distribution and topology of
pore network in the sample. Recently, similar conclusions were
reached by the authors of Ref.\,\cite{Borodich13}, who pointed out
the existing differences between materials with isolated pores and
those having more complicated topology of porous structures.

\vskip 0.05in

In this study, we examine the evolution of porous structure and
mechanical response of amorphous solids subjected to compressive
loading using molecular dynamics simulations. It will be shown that
after an isochoric quench to a temperature below the glass
transition, a variety of pore morphologies are formed, including a
random distribution of isolated voids or a connected porous network,
upon reducing average glass density. We demonstrate that under
compressive loading, the porous structure is gradually transformed
via pore coalescence and void redistribution out of the glass phase,
which results in nearly uniform density profiles at high strain.

\vskip 0.05in

The paper is organized as follows. The next section contains the
details of molecular dynamics simulation model and the deformation
procedure. Results of the numerical analysis of pore size
distributions, local density profiles, and mechanical properties of
porous glasses are presented in Sec.\,\ref{sec:Results}. The
conclusions are given in the last section.

\section{Simulation details}
\label{sec:MD_Model}


In this study, the deformation and structure of porous glassy
systems were investigated using the Kob-Andersen (KA) binary mixture
(80:20) model at a low temperature~\cite{KobAnd95}. In the KA model,
the pairwise interaction between atoms $\alpha,\beta=A,B$ is
described via the truncated Lennard-Jones (LJ) potential
\begin{equation}
V_{\alpha\beta}(r)=4\,\varepsilon_{\alpha\beta}\,\Big[\Big(\frac{\sigma_{\alpha\beta}}{r}\Big)^{12}\!-
\Big(\frac{\sigma_{\alpha\beta}}{r}\Big)^{6}\,\Big],
\label{Eq:LJ_KA}
\end{equation}
with the non-additive interaction parameters $\varepsilon_{AA}=1.0$,
$\varepsilon_{AB}=1.5$, $\varepsilon_{BB}=0.5$, $\sigma_{AB}=0.8$,
and $\sigma_{BB}=0.88$~\cite{KobAnd95}. The mass of atoms of types
$A$ and $B$ is the same, i.e., $m_{A}=m_{B}$.  The cutoff radius is
taken to be $r_{c,\,\alpha\beta}=2.5\,\sigma_{\alpha\beta}$ to
improve computational efficiency. The reduced units of length, mass,
energy, and time are defined as follows $\sigma=\sigma_{AA}$,
$m=m_{A}$, $\varepsilon=\varepsilon_{AA}$, and
$\tau=\sigma\sqrt{m/\varepsilon}$, respectively. The Newton's
equations of motion were integrated using the velocity-Verlet
algorithm~\cite{Frenkel02,Lammps} with the time step $\triangle
t_{MD}=0.005\,\tau$. The total number of atoms is $N=300\,000$. All
molecular dynamics simulations were performed using the LAMMPS
numerical code, which is designed to run efficiently in parallel
using the domain-decomposition method~\cite{Lammps}.

\vskip 0.05in

\begin{figure}[t]
\includegraphics[width=15.cm,angle=0]{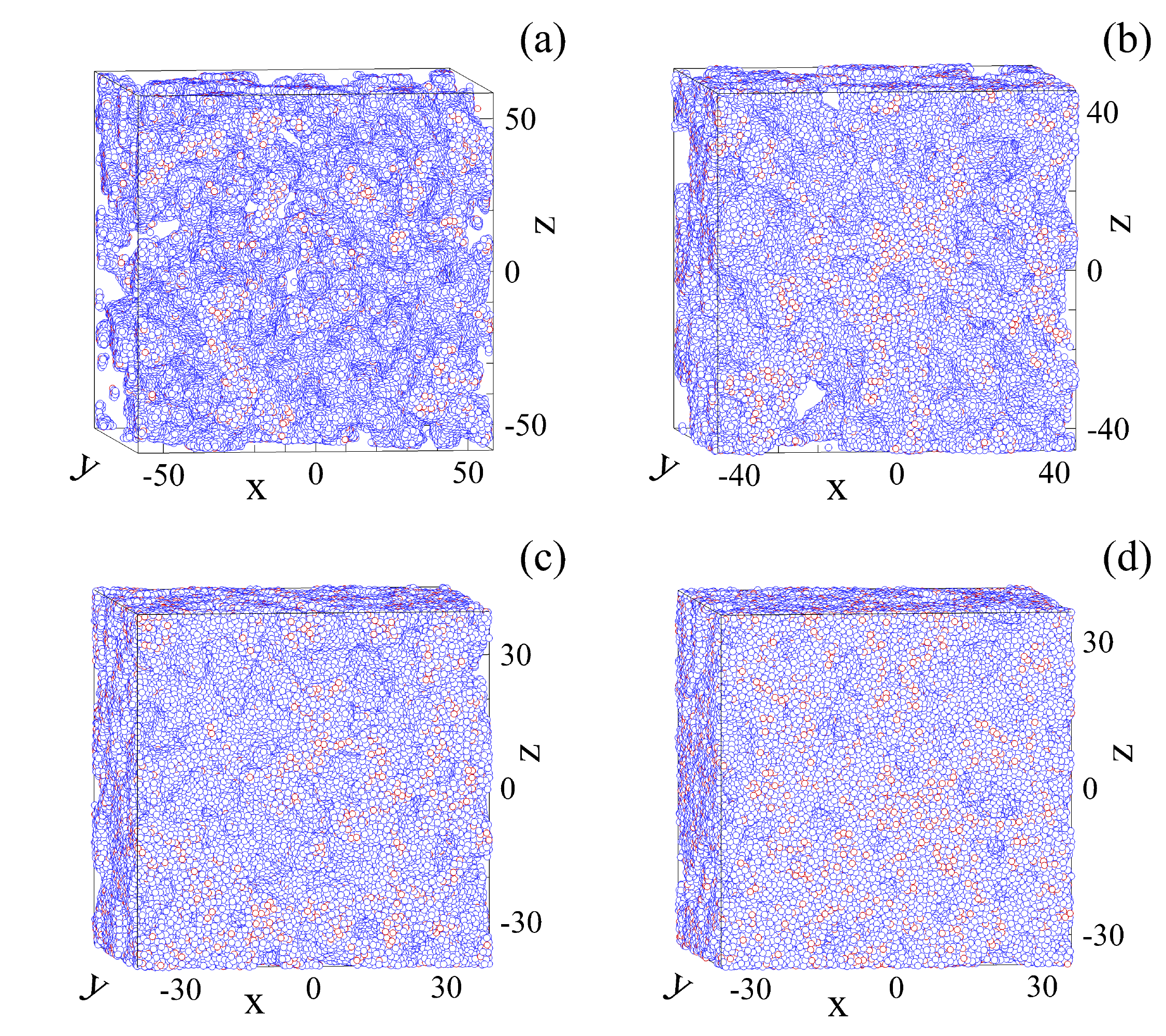}
\caption{(Color online) The representative snapshots of the porous
samples with $N=300\,000$ atoms at the temperature
$T=0.05\,\varepsilon/k_B$ for the average glass densities (a)
$\rho\sigma^{3}=0.2$, (b) $\rho\sigma^{3}=0.4$, (c)
$\rho\sigma^{3}=0.6$, and (d) $\rho\sigma^{3}=0.8$. Different atom
types are denoted by blue and red circles. Note that atoms are not
depicted to scale.}
\label{fig:snapshot_system}
\end{figure}

Following the preparation protocol used in the previous
studies~\cite{Kob11,Kob14,Priezjev17,Makeev17,Priezjev18}, the
system was first equilibrated at constant volume during $3\times
10^4\,\tau$.  At this stage, the temperature of
$1.5\,\varepsilon/k_B$, where $k_B$ is the Boltzmann constant, was
maintained by velocity rescaling.  At this temperature the binary
mixture is in the liquid phase.  To remind, the glass transition
temperature of the KA model is
$T_g\approx0.435\,\varepsilon/k_B$~\cite{KobAnd95}. The second step
involves an instantaneous quench of the system to the target
temperature of $0.05\,\varepsilon /k_{B}$ and subsequent evolution
of the system during an additional time interval of $10^{4}\,\tau$
at constant volume.   As a result of concurrent phase separation and
solidification at the low temperature, an amorphous solid with a
complex porous structure is formed, as shown, for example, in
Fig.\,\ref{fig:snapshot_system} for the average glass densities
$\rho\sigma^{3}=0.2$, $0.4$, $0.6$ and $0.8$.  In the present study,
the MD simulations were carried out in a wide range of average glass
densities, $0.2\leq\rho\sigma^{3} \leq 1.0$, and five independent
samples.

\vskip 0.05in


The compression deformation was conducted on porous glasses along
the $\hat{x}$ direction with the strain rate
$\dot{\varepsilon}_{xx}=10^{-4}\,\tau^{-1}$ at constant volume. The
maximum compressive strain is $80\,\%$, which means that the cell
size in the $\hat{x}$ direction is reduced from $L_x$ to $0.2\,L_x$
at the maximum strain. As in quiescent samples, the temperature of
$0.05\,\varepsilon /k_{B}$ was maintained via the Nos\'{e}-Hoover
thermostat~\cite{Lammps}.  During compressive deformation, the
stress tensor, potential energy, and system size were saved every
$0.5\,\tau$ for the postprocessing analysis, which was supplemented
by visual examination of consecutive atomic configurations. The data
for the elastic modulus were averaged over five independent
realizations of disorder, while representative snapshots and locally
averaged density profiles as well as pore size distributions are
reported for one sample at a given average glass density.

\section{Results}
\label{sec:Results}


The process of phase separation and concurrent solidification of a
glass-forming fluid at constant volume leads to formation of complex
porous structures in an amorphous solid~\cite{Kob11,Kob14}. With
increasing average glass density, a number of distinct morphologies
were reported at temperatures below the glass transition; namely,
disconnected droplets of the dense phase, bicontinuous structures
with increasing fraction of the solid phase, and randomly
distributed isolated pores inside the amorphous
solid~\cite{Kob11,Kob14}. Typical atomic configurations of quiescent
glassy systems considered in the present study are shown in
Fig.\,\ref{fig:snapshot_system} for the average glass densities
$\rho\sigma^{3}=0.2$, $0.4$, $0.6$ and $0.8$.   Note that pore
connectivity increases in samples with lower average glass
densities, while the structure of the solid domains remains
continuous. In our previous study, it was demonstrated that the
distribution of pore sizes in the absence of deformation is well
described by a scaling relation at small length scales and the
average glass density $\rho\sigma^{3} \lesssim 0.8$~\cite{Makeev17}.

\vskip 0.05in

\begin{figure}[t]
\includegraphics[width=12.cm,angle=0]{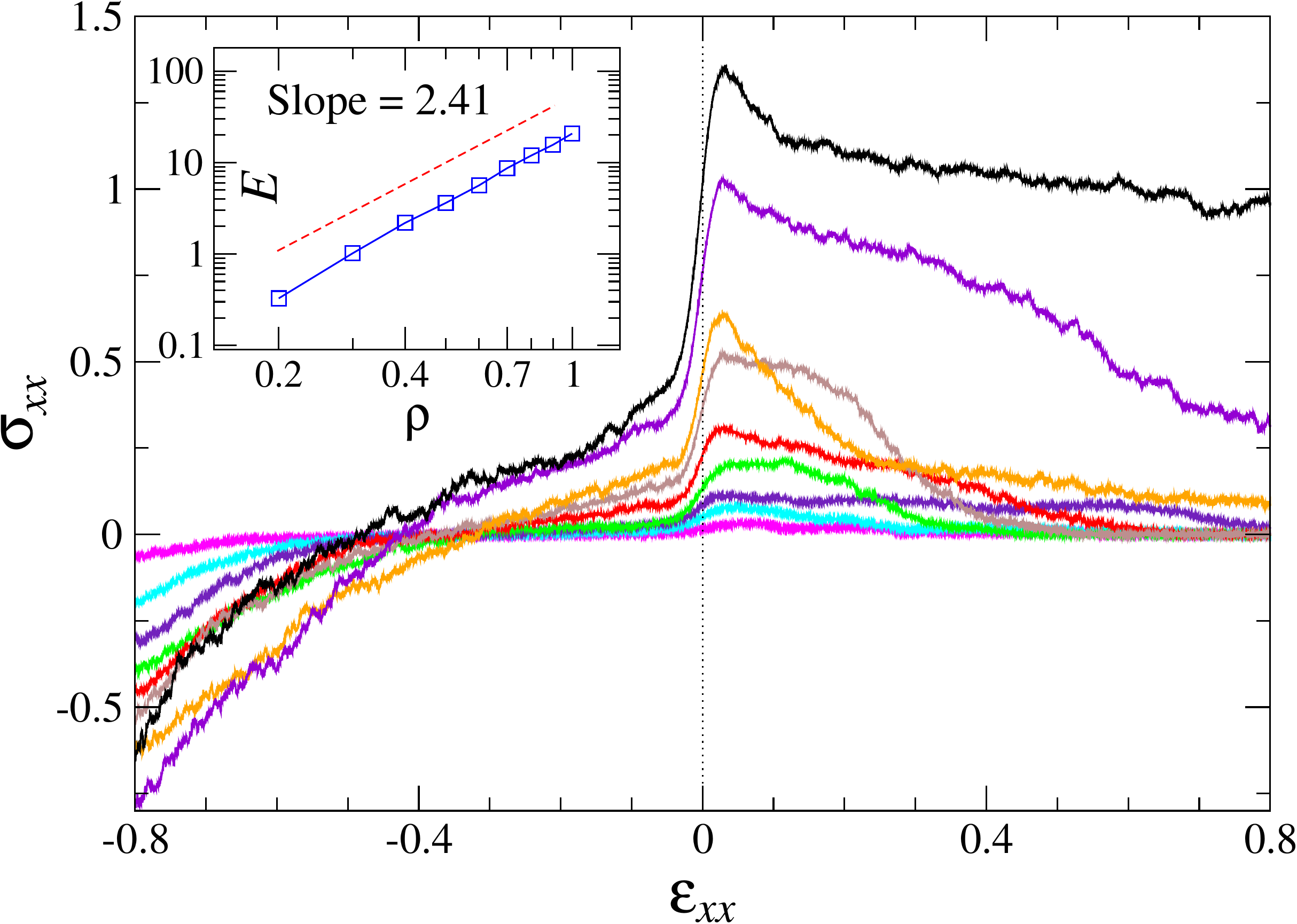}
\caption{(Color online) The strain dependence of stress
$\sigma_{xx}$ (in units of $\varepsilon\sigma^{-3}$) during
compression ($\varepsilon_{xx}<0$) and extension
($\varepsilon_{xx}>0$) with the strain rate
$\dot{\varepsilon}_{xx}=10^{-4}\,\tau^{-1}$.  The average glass
densities are $\rho\sigma^{3}=0.2$, $0.3$, $0.4$, $0.5$, $0.6$,
$0.7$, $0.8$, $0.9$ and $1.0$ (from bottom to top along the vertical
dotted line). The elastic modulus $E$ (in units of
$\varepsilon\sigma^{-3}$) as a function of $\rho\sigma^{-3}$ is
shown in the inset. The red dashed line is plotted for reference. }
\label{fig:stress_strain}
\end{figure}
\begin{figure}[t]
\includegraphics[width=15.cm,angle=0]{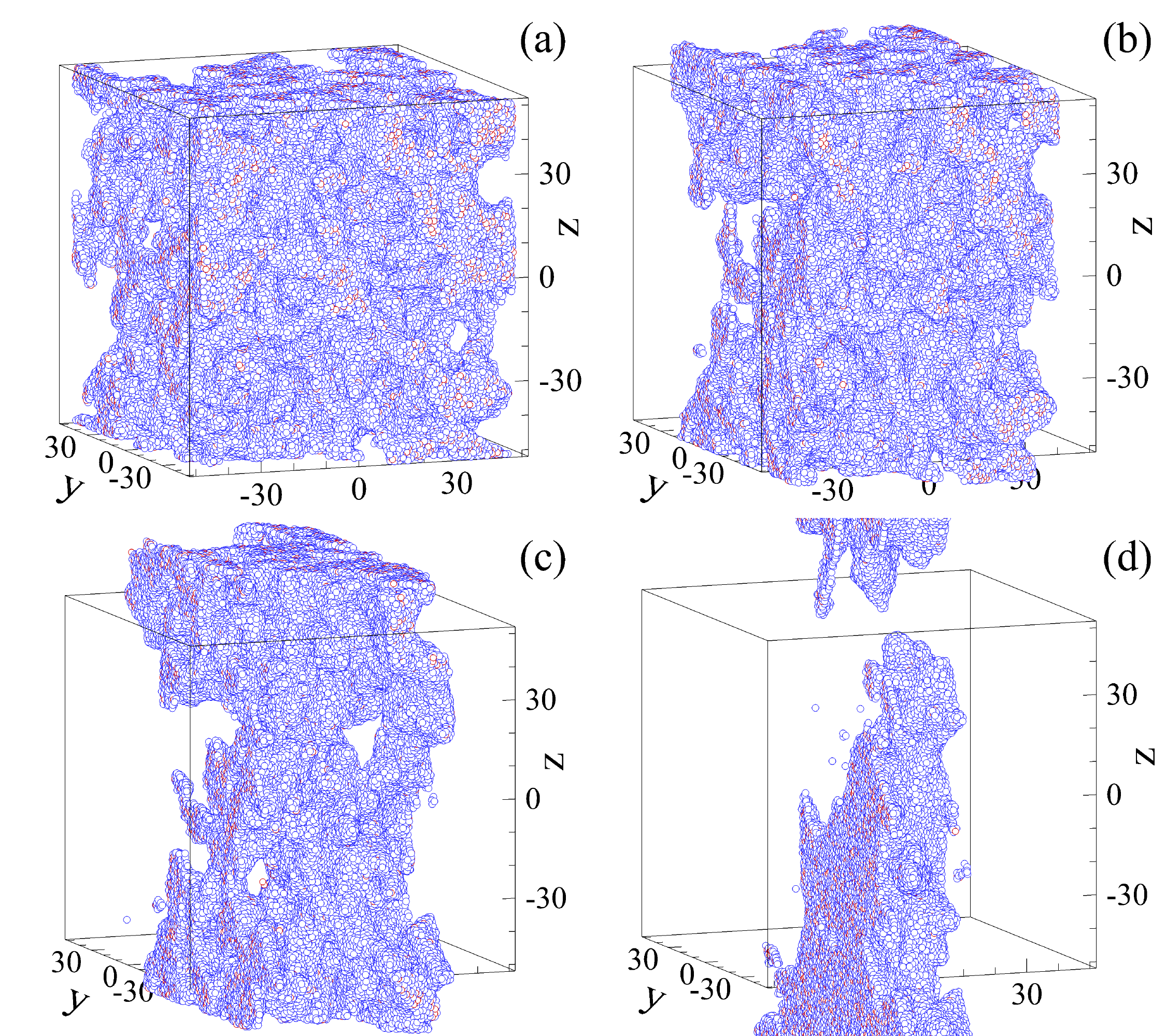}
\caption{(Color online) The snapshots of atomic configurations for
the average glass density $\rho\sigma^{3}=0.3$ and strain (a)
$\varepsilon_{xx}=0.05$, (b) $\varepsilon_{xx}=0.25$, (c)
$\varepsilon_{xx}=0.45$, and (d) $\varepsilon_{xx}=0.80$.  The
sample is compressed at constant volume with the strain rate
$\dot{\varepsilon}_{xx}=10^{-4}\,\tau^{-1}$. }
\label{fig:snapshot_strain_rho03}
\end{figure}
\begin{figure}[t]
\includegraphics[width=15.cm,angle=0]{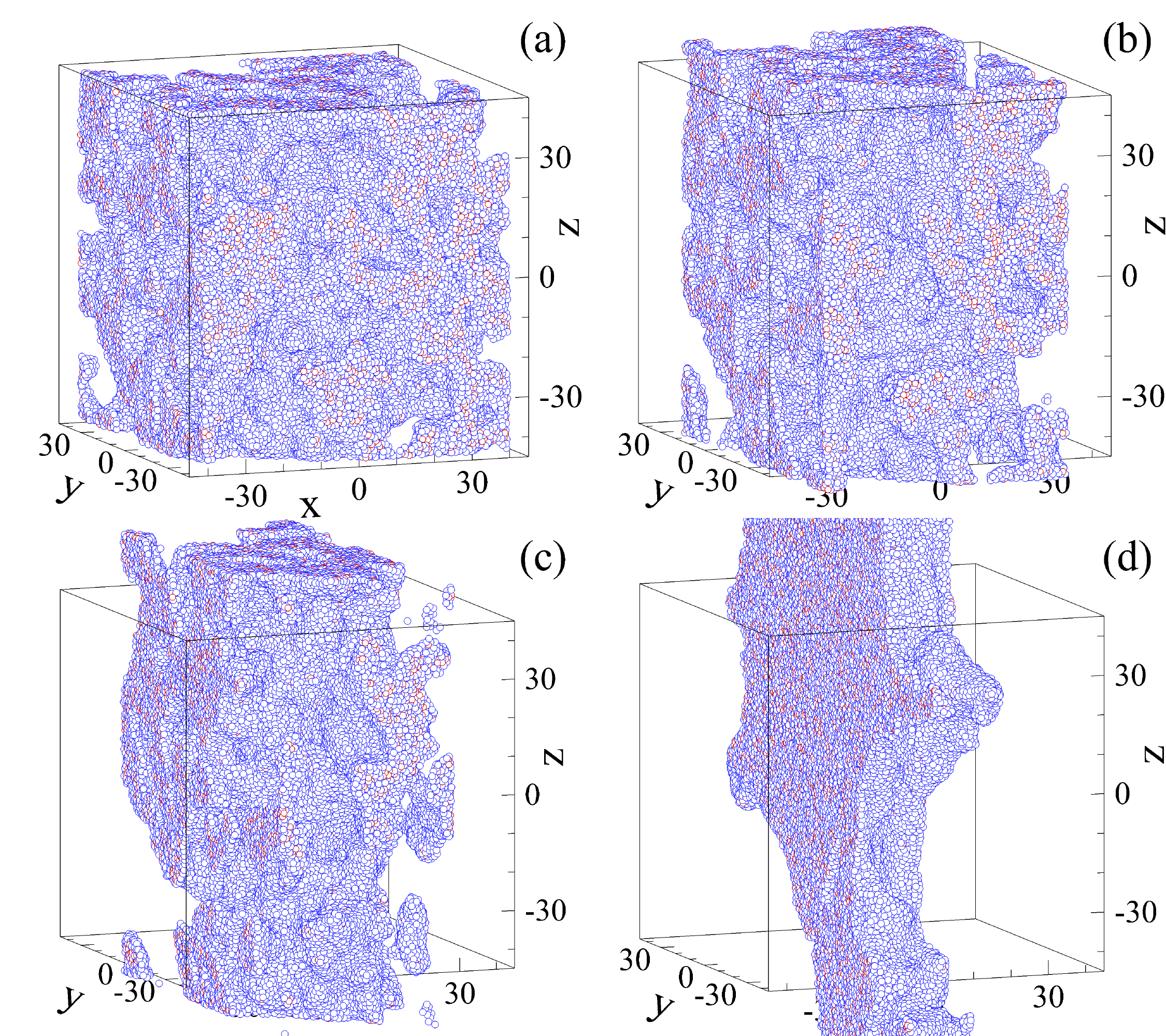}
\caption{(Color online) Four snapshots of the deformed porous glass
with the average density $\rho\sigma^{3}=0.5$ and strain (a)
$\varepsilon_{xx}=0.05$, (b) $\varepsilon_{xx}=0.25$, (c)
$\varepsilon_{xx}=0.45$, and (d) $\varepsilon_{xx}=0.80$.}
\label{fig:snapshot_strain_rho05}
\end{figure}
\begin{figure}[t]
\includegraphics[width=15.cm,angle=0]{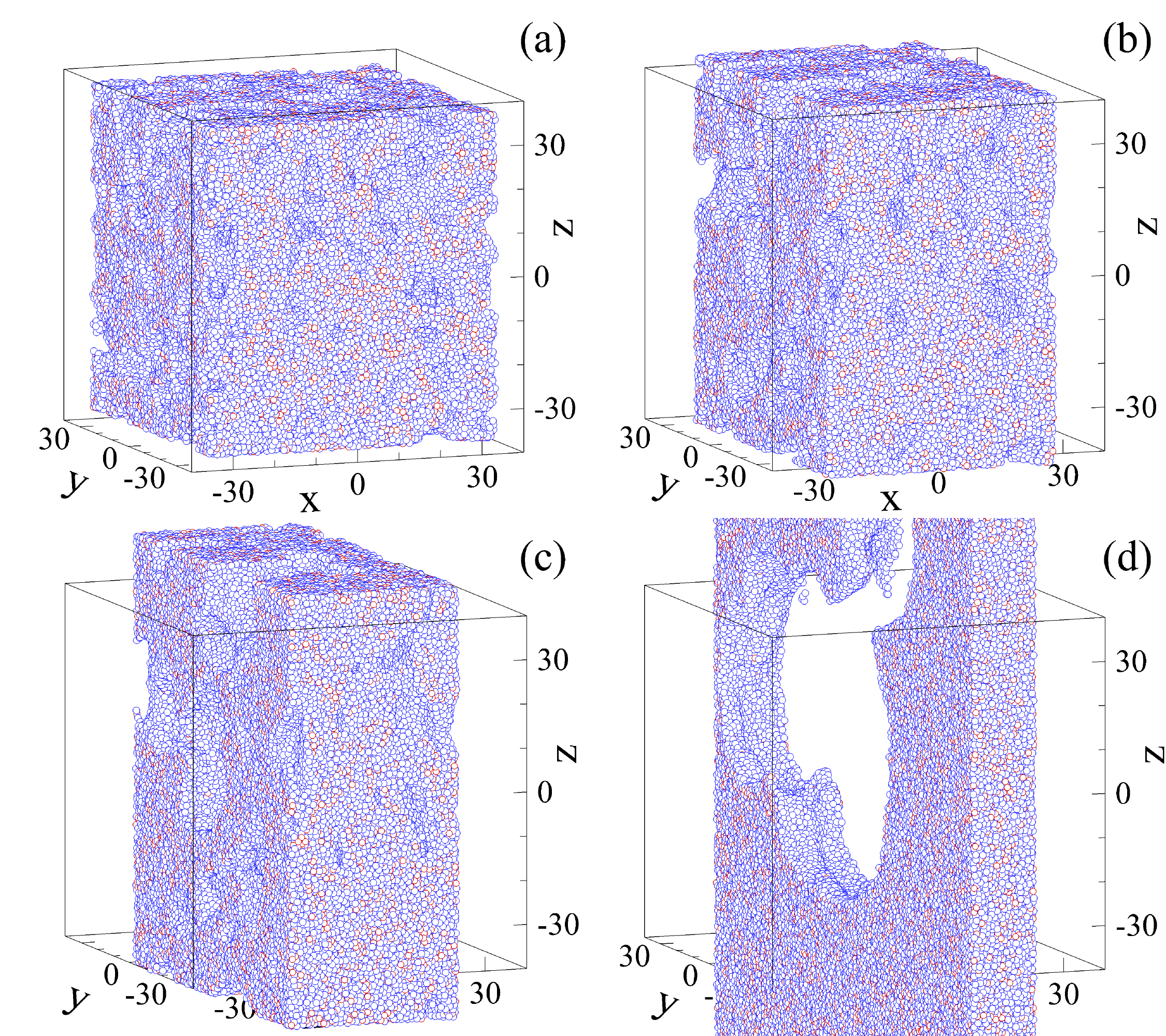}
\caption{(Color online) Snapshot images of the system configurations
at strains (a) $\varepsilon_{xx}=0.05$, (b) $\varepsilon_{xx}=0.25$,
(c) $\varepsilon_{xx}=0.45$, and (d) $\varepsilon_{xx}=0.80$. The
average glass density is $\rho\sigma^{3}=0.8$ and the strain rate is
$\dot{\varepsilon}_{xx}=10^{-4}\,\tau^{-1}$. }
\label{fig:snapshot_strain_rho08}
\end{figure}
\begin{figure}[t]
\includegraphics[width=15.cm,angle=0]{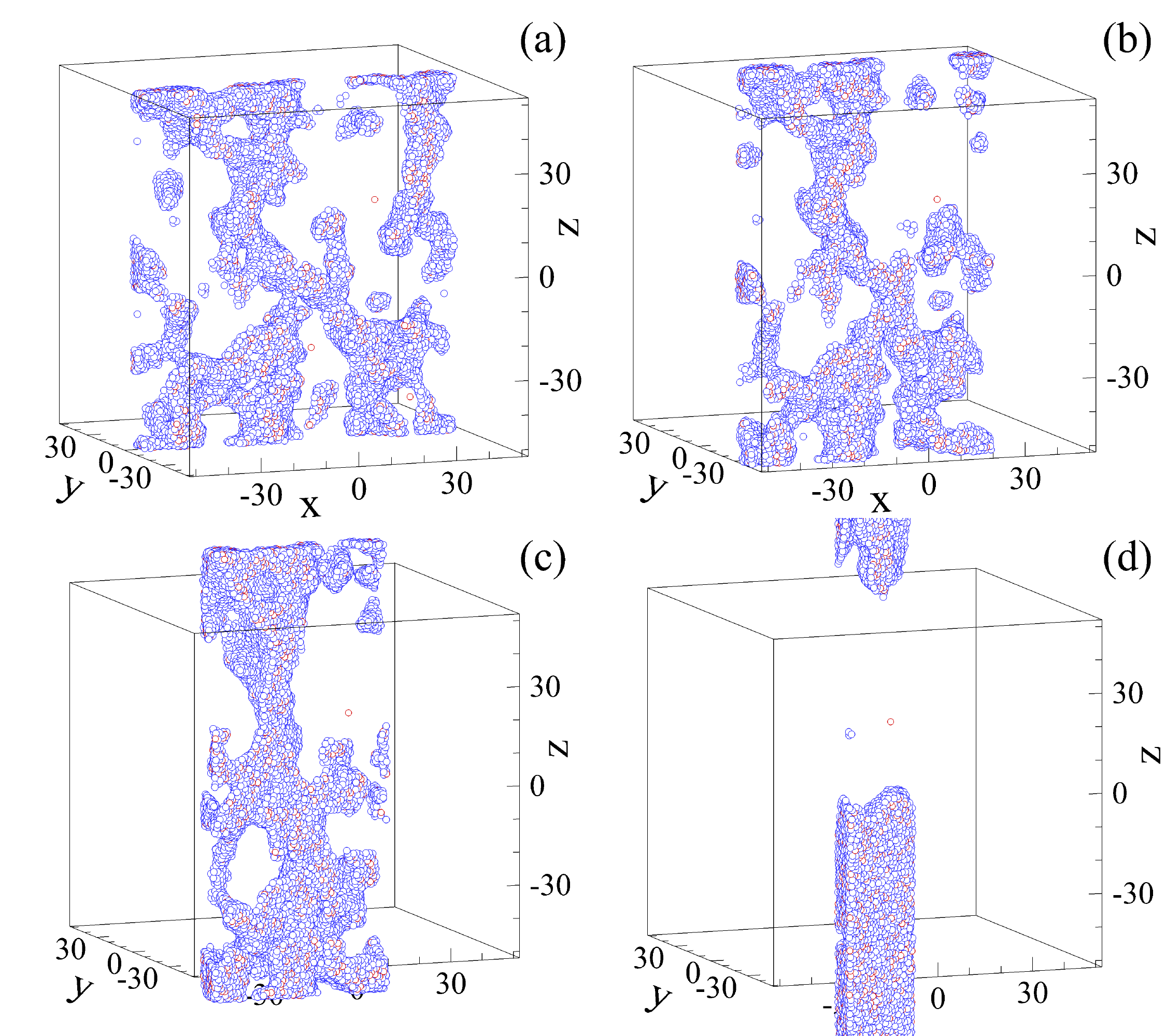}
\caption{(Color online) Reduced sets of data within a slice of
thickness $10\,\sigma$ illustrate evolution of the porous structure
in the sample with the average glass density $\rho\sigma^{3}=0.3$
and strain (a) $\varepsilon_{xx}=0.05$, (b) $\varepsilon_{xx}=0.25$,
(c) $\varepsilon_{xx}=0.45$, and (d) $\varepsilon_{xx}=0.80$.   The
same sample as in Fig.\,\ref{fig:snapshot_strain_rho03}.  }
\label{fig:snapshot_strain_rho03_slice}
\end{figure}
\begin{figure}[t]
\includegraphics[width=15.cm,angle=0]{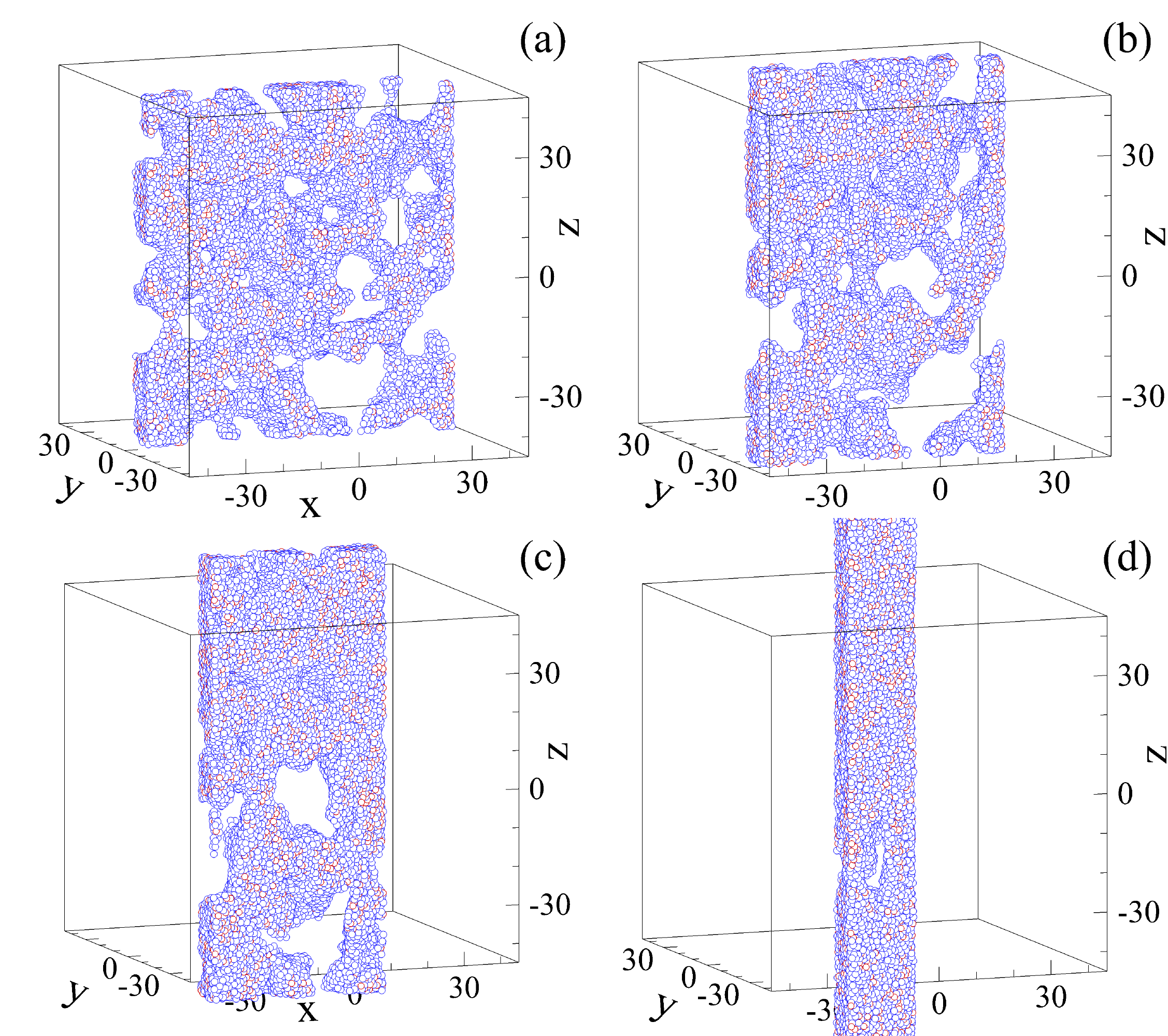}
\caption{(Color online) Atom configurations in a thin slice of
$10\,\sigma$ for the average density $\rho\sigma^{3}=0.5$ and strain
(a) $\varepsilon_{xx}=0.05$, (b) $\varepsilon_{xx}=0.25$, (c)
$\varepsilon_{xx}=0.45$, and (d) $\varepsilon_{xx}=0.80$.  The same
sample as in Fig.\,\ref{fig:snapshot_strain_rho05}. }
\label{fig:snapshot_strain_rho05_slice}
\end{figure}
\begin{figure}[t]
\includegraphics[width=15.cm,angle=0]{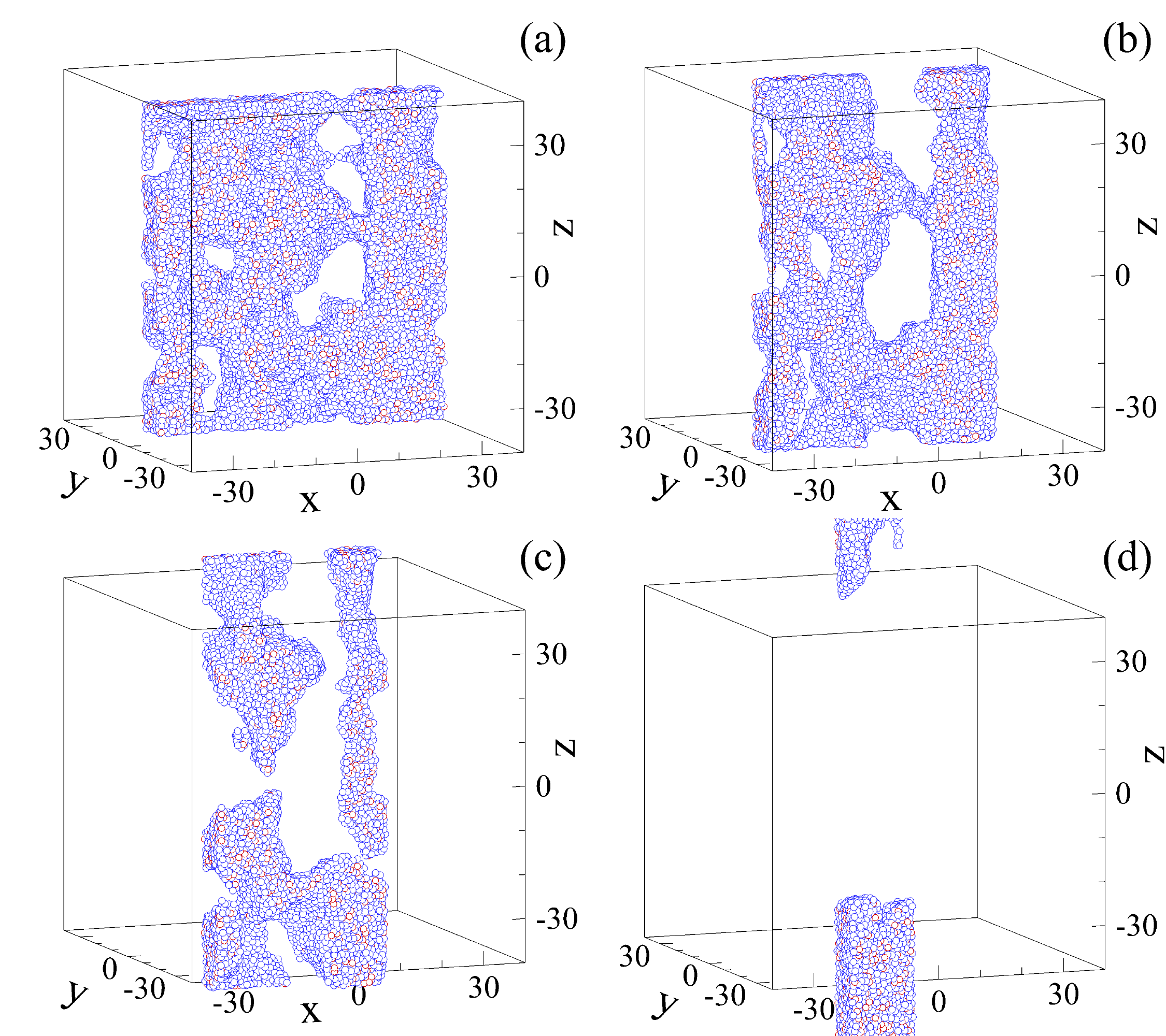}
\caption{(Color online) A series of snapshots of atom positions
within a narrow bin with thickness of $10\,\sigma$ for
$\rho\sigma^{3}=0.8$ and strain (a) $\varepsilon_{xx}=0.05$, (b)
$\varepsilon_{xx}=0.25$, (c) $\varepsilon_{xx}=0.45$, and (d)
$\varepsilon_{xx}=0.80$. The same sample as the one presented in
Fig.\,\ref{fig:snapshot_strain_rho08}. }
\label{fig:snapshot_strain_rho08_slice}
\end{figure}

Figure\,\ref{fig:stress_strain} shows the stress-strain curves
measured in one sample for each value of the average glass density
$\rho\sigma^{3}\in[0.2,1.0]$.  For completeness, the data for both
compression ($\varepsilon_{xx}<0$) and tension
($\varepsilon_{xx}>0$) deformations are reported.   Notably, the
zero strain values of stress $\sigma_{xx}$ are finite due to the
negative pressure, which arises as a result of thermal quench to the
low temperature of $0.05\,\varepsilon /k_{B}$ at constant
volume~\cite{Makeev17,Priezjev18}.   Notice that the elastic range
extends up to $|\varepsilon_{xx}|\lesssim0.04$, which is followed by
the plastic regime of deformation until the maximum strain
$|\varepsilon_{xx}|=0.8$. The elastic modulus was computed from the
slope of compressive stress $\sigma_{xx}(\varepsilon_{xx})$ at
$|\varepsilon_{xx}|\leqslant0.01$ and averaged over five independent
samples. In agreement with our previous
studies~\cite{Makeev17,Priezjev18}, the variation of the elastic
modulus as a function of the average glass density follows a
power-law increase with the exponent of $2.41$ (see inset to
Fig.\,\ref{fig:stress_strain}).   Under compressive loading, the
stress is first reduced to zero and then becomes negative, which
indicates that nearly homogeneous glass phase is accumulated in some
parts of highly strained samples, thus, providing resistance to
deformation. This effect is illustrated in consecutive systems
snapshots upon increasing strain (see
Figs.\,\ref{fig:snapshot_strain_rho03}--\ref{fig:snapshot_strain_rho08}).
In turn, the pore deformation morphologies are more clearly
visualized in narrow slices across the systems shown in
Figs.\,\ref{fig:snapshot_strain_rho03_slice}--\ref{fig:snapshot_strain_rho08_slice}
for the average glass densities $\rho\sigma^{3}=0.3$, $0.5$ and
$0.8$.

\vskip 0.05in


In this work, the pore size distribution (PSD) functions are
obtained using the open-source software
ZEO++~\cite{Haranczyk12c,Haranczyk12,Haranczyk17}. The approach is
based upon a Voronoi network representation of the accessible void
space. Specifically, Voronoi network consists of nodes and edges
mapping the space around atoms in the system. Each node and edge
contain information on the distances to the nearest atoms; the
distances correspond to the radii of the \textit{largest} probe of a
spherical shape that can move along the edge without intersecting
any atom.  The probe-accessible regions are found via the modified
Dijkstra shortest-path algorithm~\cite{Dijkstra59}. Within this
framework, the probe-accessible regions of the periodic Voronoi
network are represented by a sub-graph. Thereby, an atomic structure
can be converted into a periodic graph-representation of the void
space for a given radius of the probe.

\vskip 0.05in

\begin{figure}[t]
\includegraphics[width=12.cm,angle=0]{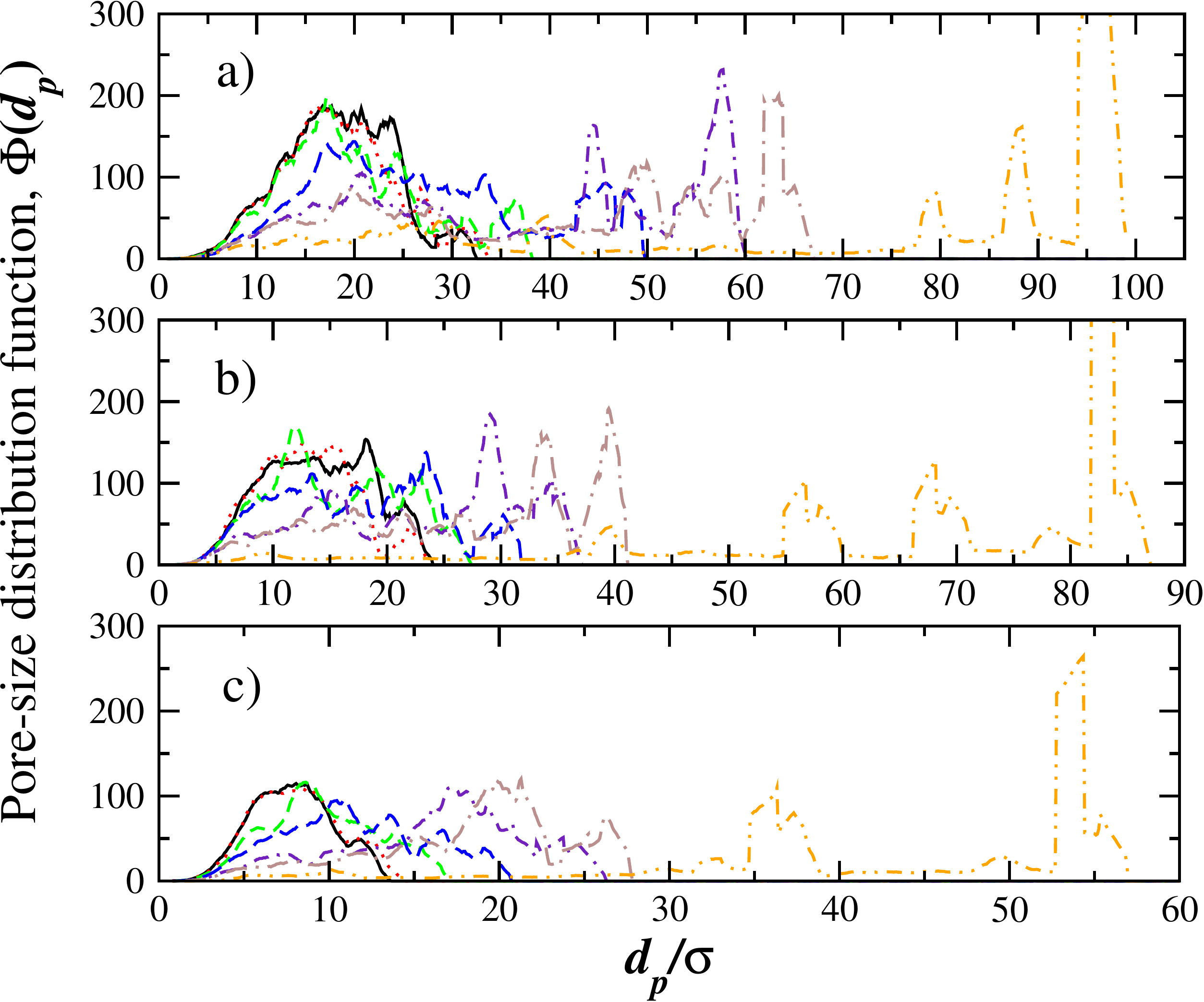}
\caption{(Color online) The pore size distribution functions for the
cases (a) $\rho\sigma^{3}=0.3$, (b) $\rho\sigma^{3}=0.5$, and (c)
$\rho\sigma^{3}=0.8$.  The colorcode for different curves is as
follow: solid black ($\varepsilon_{xx}=0.0$), dotted red
($\varepsilon_{xx}=0.05$), dashed green ($\varepsilon_{xx}=0.15$),
dashed blue ($\varepsilon_{xx}=0.25$), dash-dotted indigo
($\varepsilon_{xx}=0.45$), dash-dotted brown
($\varepsilon_{xx}=0.50$), and double-dot-dashed orange
($\varepsilon_{xx}=0.75$).}
\label{fig:pore_size_dist}
\end{figure}

The pore size distribution functions,  $\Phi(d_p)$, are shown in
Fig.\,\ref{fig:pore_size_dist}. Here, we present our results for the
cases of average densities $\rho\sigma^{3}=0.3$, $0.5$ and $0.8$.
The PSD functions in quiescent samples for the same set of
$\rho\sigma^{3}$ values have been studied in Ref.\,\cite{Makeev17}.
It was shown that PSDs are narrow at high glass densities and they
become broader as the average glass density decreases; various
features of PSDs were discussed in the same reference. In the past,
we have also investigated evolutions of PSDs in porous glasses
undergoing shear~\cite{Priezjev17} and tensile~\cite{Priezjev18}
loadings. This allows for a comparative analysis of the data,
obtained on different types of mechanical loading.

\vskip 0.05in


First, we found that the general behavior of the PSDs under
compression is similar to the cases of shearing and tension at small
and intermediate strain deformation. Indeed, when strain is small,
the shape of PSD curves shown in Fig.\,\ref{fig:pore_size_dist}
remains largely unaffected, and the widening of the PSDs is small.
With increasing strain, the PSDs widen significantly and gradually
start to develop a double-peak profile. The same type of behavior
has been observed in the systems undergoing structural evolution
under shear and tension~\cite{Priezjev17,Priezjev18}. In the cases
under consideration, the magnitudes of PSDs decrease drastically
with strain in the regions of small values of $d_p$. At the same
time, the magnitudes of peaks, newly developed at large values of
$d_p$, increase (see Fig.\,\ref{fig:pore_size_dist}). This is
consistent with the results reported for porous glasses under shear
and tension~\cite{Priezjev17,Priezjev18}. However, compression above
the threshold value of  $\approx\!0.7$ leads to a nearly complete
separation of the porous glass into solid domain and void space. At
densities $\rho\sigma^{3}=0.3$ and $0.5$, the separation leads to a
formation of the void space region with linear dimension exceeding
that of a half of the system size in the direction perpendicular to
the loading axis. According to our analysis, a substantial
densification takes place in the solid domains, with the maximum of
the density profile along the $\hat{y}$ direction being more than
twice the average density of the system (not shown).

\vskip 0.05in


Second, previously we found that, in the case of tension, the system
with density $\rho\sigma^{3}=0.8$ shows anomalously large peak,
corresponding to large-size pore developed in the
system~\cite{Priezjev18}.  Here, again, the same type of behavior is
observed. The system undergoes a rapid separation into high-density
solid material (glass) and large voids. Note that the finite-size
effects may interfere with the process of void growth, when a pore
diameter becomes close the system's characteristic dimension.
Therefore, one should exercise some caution in interpreting the
data, when dimensions of the pores approach those of the simulation
box. The growth of large-size voids is accompanied by significant
decrease in the number of pores having dimension characteristic for
unperturbed samples. Indeed, the small-size pores nearly disappear,
when $\varepsilon_{xx}$ exceeds $0.5$ in the system with
$\rho\sigma^{3}=0.8$ shown in Fig.\,\ref{fig:pore_size_dist}.
Altogether, the general conclusions are similar to the cases of
shearing and tension~\cite{Priezjev17,Priezjev18}. Those can be
summarized as follows. Compressive deformation leads to multiple
structural transitions, characterized by gradual evolution from a
number of small-size compact pores to a configuration with one or
two dominant pores.

\vskip 0.05in


A temporal picture of material rearrangement under compression can
be unveiled by considering spatially-resolved (coarse-grained)
density profiles at a chosen sets of compressive strain magnitudes.
Similar to our study of porous glass response to
tension~\cite{Priezjev18}, here we consider spatially-resolved
average density, computed along the direction of compressive
loading, $\langle \rho \rangle_{s}(x)$.  The quantity is defined as
the number of atoms located in a bin with thickness $b\approx\sigma$
along the $\hat{x}$ direction (the direction of the loading),
divided by the volume of the bin: $b\,L_yL_z$, where, $L_y$ and
$L_z$ are the box sizes in the two Cartesian directions
perpendicular to the loading direction. In Ref.\,\cite{Priezjev18},
we found that the failure under tension occurs in large-scale,
low-density regions. In other words, the location of the failure is
correlated to the low-density regions of large spatial
extents~\cite{Priezjev18}.

\vskip 0.05in

\begin{figure}[t]
\includegraphics[width=12.cm,angle=0]{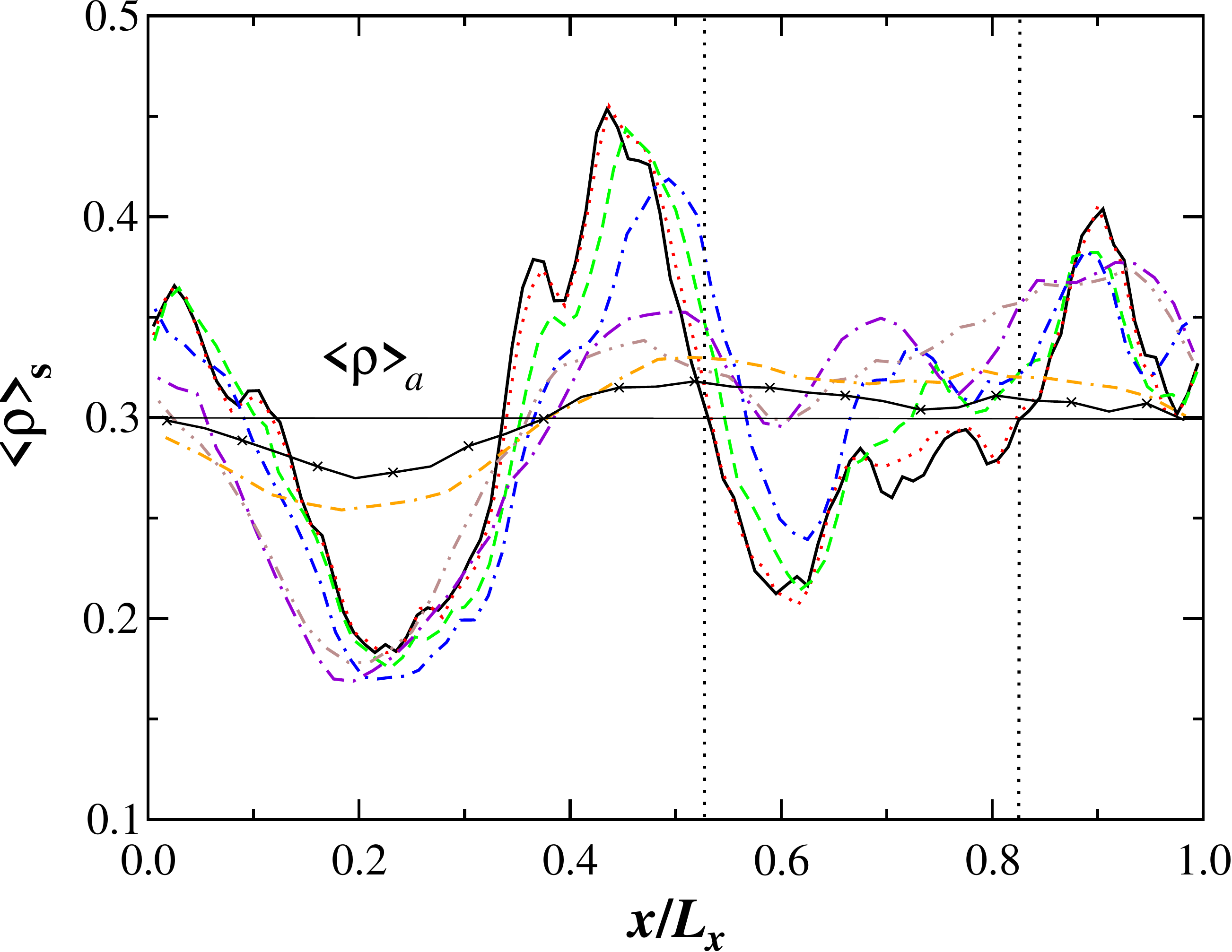}
\caption{(Color online) The local density profiles $\langle \rho
\rangle_{s}(x)$ (in units of $\sigma^{-3}$) for the values of strain
$\varepsilon_{xx}=0.0$ (solid black), $0.05$ (dotted red), $0.15$
(dashed green), $0.25$ (dash-dotted blue), $0.45$ (dash-dotted
violet), $0.5$ (dash-dotted brown), $0.75$ (dash-dotted orange), and
$0.8$ (black curve with crosses). The horizontal line denotes the
average glass density $\rho\sigma^{3}=0.3$.  The two vertical dotted
lines indicate the region with reduced density. }
\label{fig:den_prof_rho03}
\end{figure}
\begin{figure}[t]
\includegraphics[width=12.cm,angle=0]{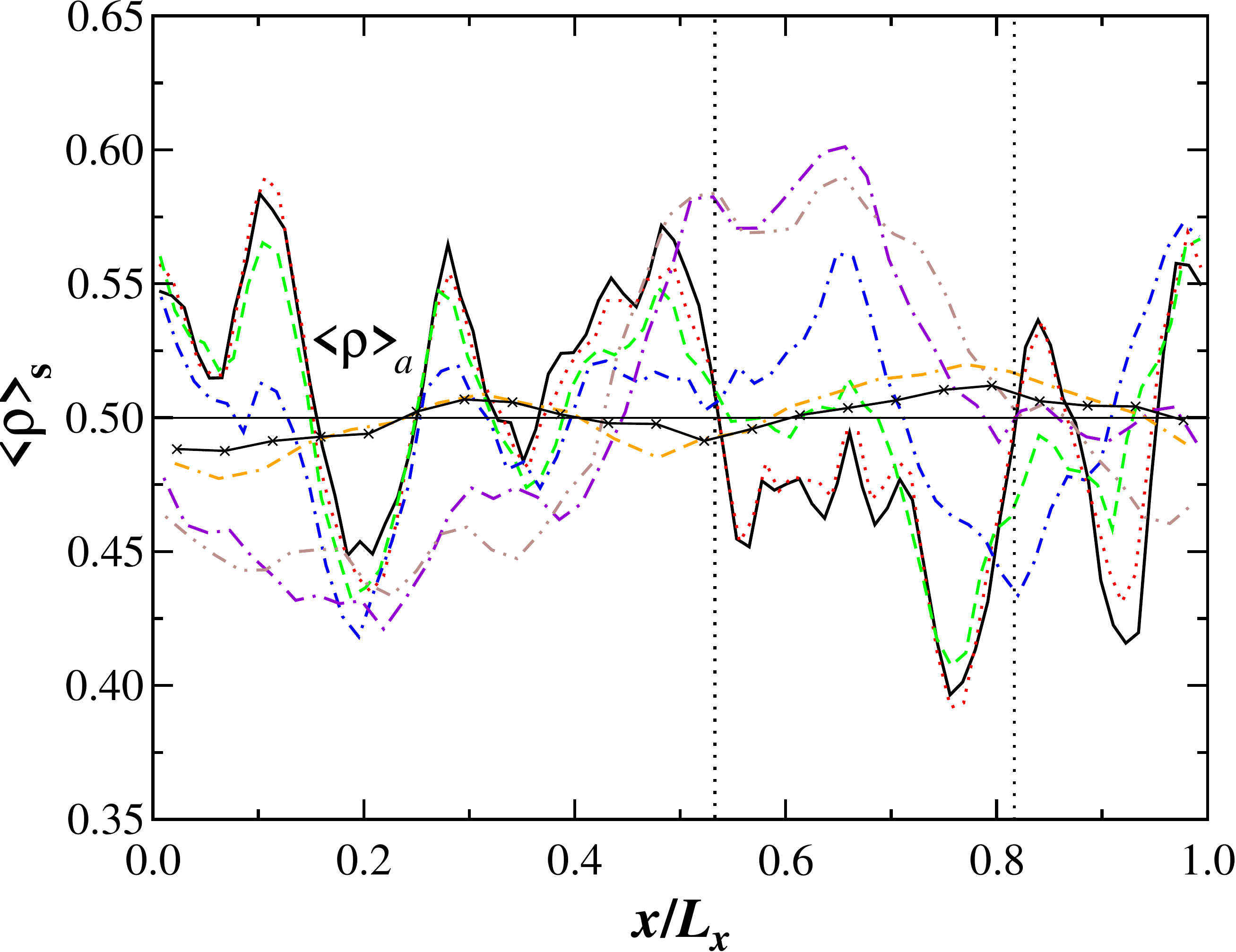}
\caption{(Color online) Spatially-resolved density profiles $\langle
\rho \rangle_{s}(x)$ (in units of $\sigma^{-3}$) for the same values
of compressive strain as in Fig.\,\ref{fig:den_prof_rho03}. The
average glass density $\rho\sigma^{3}=0.5$ is shown by the
horizontal line. The same colorcode as in
Fig.\,\ref{fig:den_prof_rho03}.  }
\label{fig:den_prof_rho05}
\end{figure}
\begin{figure}[t]
\includegraphics[width=12.cm,angle=0]{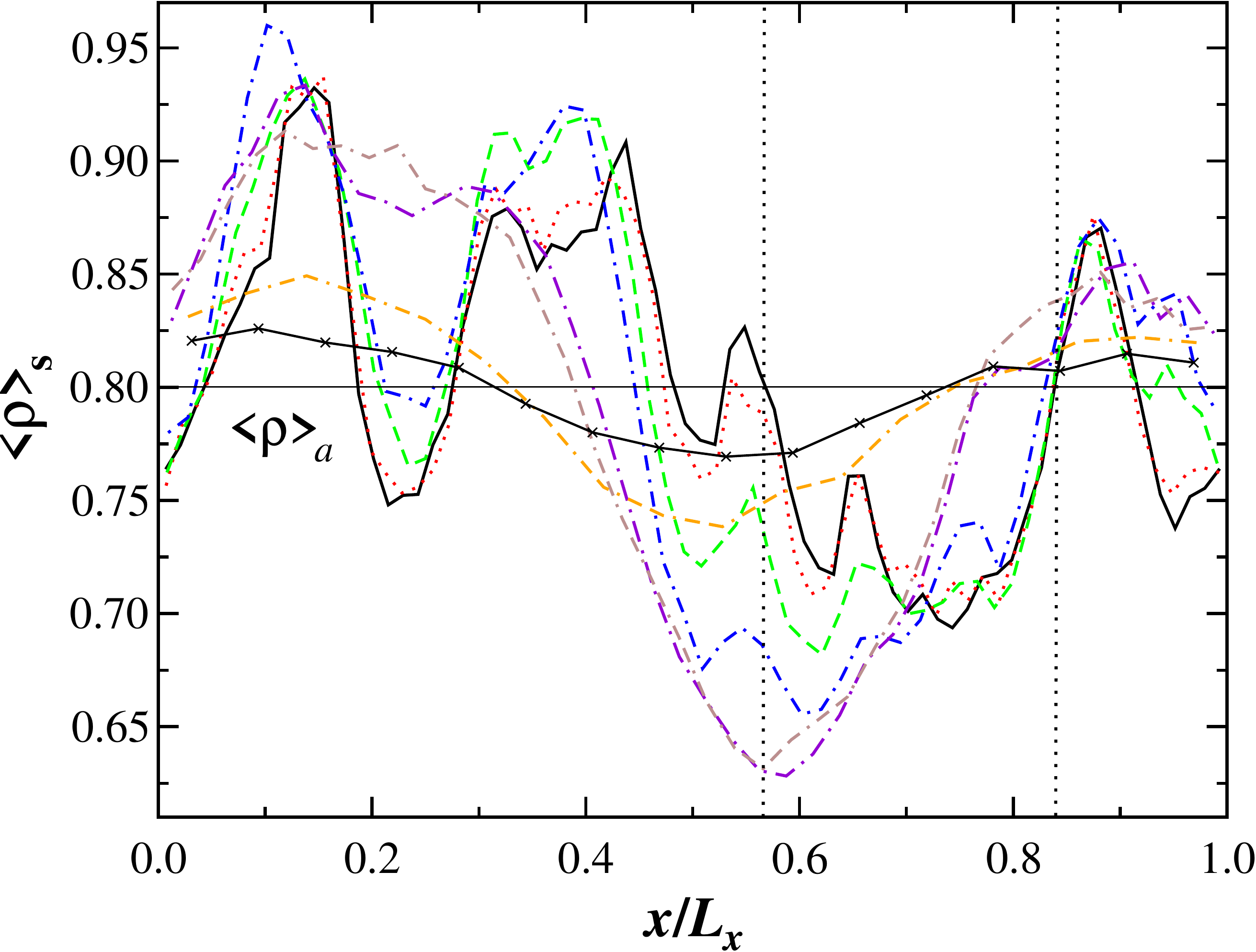}
\caption{(Color online) The density profiles $\langle \rho
\rangle_{s}(x)$ (in units of $\sigma^{-3}$) of the solid phase for
$\rho\sigma^{3}=0.8$ and selected values of compressive strain.  The
colors and strain values are the same as in
Figs.\,\ref{fig:den_prof_rho03}.}
\label{fig:den_prof_rho08}
\end{figure}

As shown below, in the case of compressive loading, there also
exists a number of notable events pertained to the structural
evolution. Correspondingly, in
Figs.\,\ref{fig:den_prof_rho03}--\ref{fig:den_prof_rho08}, we mark
the low-density regions by dashed vertical lines. As follows from
Figs.\,\ref{fig:den_prof_rho03}--\ref{fig:den_prof_rho08}, two
different types of behavior can be discriminated depending on the
average density of porous glasses. Next, we consider the two types
separately.  In the low- and intermediate-density systems
($\rho\sigma^{3}=0.3$ and $0.5$), the compression induces a
densification of the regions with initially low local densities,
while the density shows some decrease in the domains, where
$\rho\sigma^{3}$ was above the average value before loading. In a
sense, the effect can be described as a gradual rearrangement of the
glassy material, such that the local density along the direction of
compression is homogeneous and equals to the average density. This
behavior is characteristic for small and intermediate values of
compressive strains.   As the strain magnitude approaches
$\varepsilon_{xx}=0.7$, an abrupt homogenization occurs, with the
average density rapidly approaching the average value, i.e.,
$\rho\sigma^{3}=0.3$ and $0.5$ (see Figs.\,\ref{fig:den_prof_rho03}
and \ref{fig:den_prof_rho05}).

\vskip 0.05in


The behavior is markedly different for systems with
$\rho\sigma^{3}=0.8$ shown in Fig.\,\ref{fig:den_prof_rho08}. In
this case, the zone of low-density (marked by vertical dashed lines)
does not undergo densification. To the contrary, a density dip
develops within the zone, moves towards one border of the
low-density zone and gets localized at this border at intermediate
values of $\varepsilon_{xx}$.   At high strain magnitudes, an
apparent pore closures take place in the whole sample, the density
is close to homogeneous, and its value approaches the average
density, $\rho\sigma^{3}=0.8$.  We would like to reiterate that the
porous glasses in our study were formed in confined environment.
Correspondingly, we perform the compressive loading tests at
constant volume. Therefore, the systems are different from the
seemingly equivalent set-ups, widely used in the studies of
shock-wave propagation through porous media.  In the case of
shock-wave propagation, a density increase in the after-shock
regions takes place, with the shocked material density being in
excess of its thermodynamic equilibrium
value~\cite{Simons82,Trunin89,Makeev09,Soulard15}. In the case,
considered herein, we rather observe a nearly complete separation of
bimodal systems in solid domains and void, with density of these
solid domains gradually approaching its equilibrium value for the
void-free state ($\rho\sigma^{3}=1.25$~\cite{{Stillinger11}}). Since
the volume is conserved (unlike the case of shock-compression), the
average density approaches its corresponding value throughout the
samples. The systems we consider can be of relevance to the problems
related to gas absorption or flow of gas (fluid) in nano-porous
materials, where stresses can arise, for example, from
adsorbates~\cite{Huber17}. The problem of shock-waves in the system
under consideration will be addressed in a future work.

\section{Conclusions}

In summary, we reported the results of a molecular dynamics study
aimed at understanding the influence of pore and glass structures on
compressive loading at constant volume and low temperature. For a
given average glass density, the porous samples were produced at
constant volume through kinetically arrested liquid-gas spinodal
decomposition of a glass-forming system at a temperature well below
the glass transition. The resulting structures are characterized by
a connected porous network at lower average glass densities and a
collection of randomly distributed compact pores at higher average
glass densities. In addition, in a wide range of average glass
densities, the pressure in quiescent samples is negative due to a
distribution of built-in tensile stresses in the solid domains. We
found that upon compressive loading, the axial stress is first
reduced to zero, thus releasing built-in stresses in the system, and
then it becomes negative due to accumulation of the homogeneous
glass at high strain. In agreement with our previous results on
tensile and shear deformation, the compressive elastic modulus
increases as a power-law function of the average glass density.
Finally, the numerical analysis of density profiles and pore size
distributions during compressive loading showed that porous
structures undergo significant deformation and expulsion from the
glass phase leading to effective separation of empty regions from
the homogeneous solid domains.

\section*{Acknowledgments}

Financial support from the National Science Foundation (CNS-1531923)
is gratefully acknowledged.  The molecular dynamics simulations were
performed using the LAMMPS numerical code developed at Sandia
National Laboratories~\cite{Lammps}. The study has been in part
funded by the Russian Academic Excellence Project `5-100'.
Computational work in support of this research was performed at
Michigan State University's High Performance Computing Facility and
the Ohio Supercomputer Center.


%

%


%

%

%

%

%

%

%

%

%

%

\bibliographystyle{prsty}

\end{document}